# A COMPUTATIONAL MODEL REVEALING THE EFFECT OF DOPAMINE ON ACTION SELECTION

Neslihan S. Sengor, Ozkan Karabacak

Istanbul Technical University, Faculty of Electrical and Electronic Engineering, Electronics Engineering Department, Maslak, TR-34469, Istanbul, Turkey

**Contacting Author:**

Ozkan Karabacak

Istanbul Technical University, Faculty of

Electrical and Electronic Engineering,

Electronics Engineering Department,

Maslak, TR-34469, Istanbul, Turkey

Phone: +90 212 2853617

Fax: +90 212 2853679

E-mail: karabacak@ehb.itu.edu.tr

# A COMPUTATIONAL MODEL REVEALING THE EFFECT OF DOPAMINE ON ACTION SELECTION

## Abstract

In order to reveal the effect of nigrostriatal dopamine system on action selection, first a computational model of the cortex-basal ganglia-thalamus loop is proposed and based on this model a simple compound model realizing the Stroop effect is established. Even though Stroop task is mostly used to examine selective attention, the main objective of this work is to investigate the effect of action selection on Stroop task. The computational model of the cortex-basal ganglia-thalamus loop is a non-linear dynamical system which is not only capable of revealing the action selection property of basal ganglia but also capable of modelling the effect of dopamine on action selection. While the interpretation of action selection is based on the solutions of the non-linear dynamical system, the effect of dopamine is modelled by a parameter of the model. The inhibiting effect of dopamine on the habitual behaviour which corresponds to word reading in Stroop task and letting the novel one occur corresponding to colour naming is investigated using the compound computational model established in this work.

**List of Symbols and Abbreviations:**

| | |
|---|---|
| **AI** | artificial intelligence |
| **BG** | basal ganglia |
| **C** | cortex |
| **GPe** | globus pallidus externia |
| **GPi** | globus pallidus internia |
| **HD** | Huntington's disease |
| **PD** | Parkinson's disease |
| **STN** | subthalamic nucleus |
| **SNr** | substantia nigra pars reticulata |
| **TH** | thalamus |
| **TSSG** | temporal sequence storage and generation |

# 1 Introduction

Recently, computational modelling of cognitive processes began to draw more attention from different disciplines [Cohen D., & Servan-Schreiber D., (1992), Durstewitz D., &Seamans J.K., (2002), Behrendt R.P., (2004), Doya K., (2000)]. Due to the diversity of disciplines dealing with modelling cognitive processes, there are different computational approaches. Some approach the modelling problem to get a biophysically realistic model and thus begin to set models of cells at physiological level [Durstewitz D., & Seamans J.K., (2002), Ashby F.G., & Casale M.B., (2003)]. This requires exact knowledge of the facts going on at each cell, which cannot be easily fulfilled. Another drawback of this approach, which is also more important for computational modelling, is that any potential model of a simple behaviour would be so complex that realizing it could not be possible [Churchland P.S., (1988), Grossberg S. (1980)]. Others approach from a different perspective and use symbolic artificial intelligence (AI) techniques or production systems [Norman D.A., & Schallice T., (1986), Goel V., Pullara, S.D., & Grafman J., (2001)]. These also have drawbacks; even though the model is now capable of realizing the behaviour, it is far away from explaining which neural substrates affect each other and how they accomplish the involved cognitive process.

The aim of modelling has to be more than to get a system behaving as expected, the model also has to reveal what is silently going on during the observed behaviour. Another approach focuses on this aspect of computational modelling and it deals with modelling at behavioural level. These models are not as complex as the biophysically realistic models and not as far away from the neural validity as symbolic AI techniques and production systems. Neural substrates related with the involved behaviour and their interrelations are considered while modelling at behavioural level, thus the model is capable of explaining how the neural substrates provoke the behaviour without dealing with structures at physiological level [Braver T.S., & Cohen D.C., (2000), Taylor J.G., & Taylor N.R., (2000), Taylor N. R., &

Taylor J.G., (2000), Gurney K., Prescott T.J., & Redgrave P., (2001a,b), Suri R.E., Bargas J., & Arbib M.A. (2000), Holyrod C.B., & Coles M.G.H. (2002)]. Modelling at behavioural level is not only beneficial at explaining the whys of cognitive processes but this approach also provides tools for investigating the reasons behind the abnormal behaviour. This attribute of behavioural models makes them especially important in pharmacological studies [Ashby, F.G., & Casale, M.B., (2003)]. As animal models are not sufficient, and behavioural models provide more flexible applications, they are advantageous over models used nowadays in pharmacology. Still another advantage of these models is they inspire new approaches in engineering applications as designing robots and intelligent systems and even motivate new ideas for economical systems [Doya, K., (2000), Cohen J.D., (2005)].

Our aim is to demonstrate that a functional model, which may be quite simple compared to the biological system to be modelled, could still be informative in explaining the relations between neural substrates and effect of neurotransmitters. The observed physical reality can be modelled as a non-linear dynamical system and such a model not only provides a tool to understand the relations between neural substrates and the observed behaviour but it also provides a prospective look and helps to see not yet thought relations and explanations. These have been basics of scientific work especially in physics where mathematics and physics developed together. This approach could also be used in life sciences even in most unexpected subjects as modelling behaviours. To reveal the accomplishment of our aim, we will present a functional model of cortex, basal ganglia and thalamus loop realizing action selection function, especially considering the effect of the neurotransmitter dopamine in basal ganglia and the proposed model is at behavioural level.

Basal ganglia have been the most thoroughly studied neural structure and thus numerous computational models related to this structure have been proposed [Amos,A., (2000), Taylor J.G., & Taylor N.R., (2000), Gurney K., Prescott T.J., & Redgrave P., (2001a,), Suri R.E.,

Bargas J., & Arbib M.A. (2000), Holyrod C.B., & Coles M.G.H. (2002)]. The role of basal ganglia in cognitive processes are now more appreciated, this is an inspiring progress, as they are once considered just by their duty in motor control. As proposed in [Alexander, G.E., & Crutcher M.D., (1990)], the activity of basal ganglia is not the only factor in establishing cognitive processes, basal ganglia with cortex and thalamus forms basal ganglia-thalamocortical circuits such that for different processes different substructures of cortex, basal ganglia and thalamus are in action. In [Alexander, G.E., & Crutcher M.D., DeLong M.R., (1990)] five such circuits have been introduced and it is proposed that some of these circuits have roles in cognitive processes. In each circuit basal ganglia have a role and especially their action selection property has impacts on different cognitive processes [Mink, J.W., (1996), Kropotov J.D., & Etlinger S.C., (1999), Behrendt R. P., (2004), Graybiel, A.M., (2004)]. There have been various attempts to model this property of the basal ganglia [Fukai T., (1999), Gurney K., Prescott T.J., & Redgrave P., (2001a,b)] and in [Kropotov J.D., & Etlinger S.C., (1999)] it is even stated that "the ability to select an appropriate action at a given time interval determines cognitive functions of the human brain".

The main objective of this work is to set a model of cortex, basal ganglia, thalamus (C-BG-TH) loop, which would be prone to explain action selection property of the basal ganglia-thalamocortical circuits. In the meantime, the model is also capable of demonstrating the effect of nigrostriatal dopamine system on action selection. The nigrostriatal dopamine system has been mostly considered effective in motor control, but there are evidences showing that nigrostriatal dopamine system also affects cognitive functions [Graybiel A.M., (1990), Graybiel, A.M., (1995), Horvitz, J.C. (2000), Djurfeldt M., Ekeberg Ö., &Graybiel A.M., (2001), Monchi, O., Petrides M., Doyon, J., Postuma, R.B., Worsley, K., & Dagher, (2004)]. Furthermore, the impact of action selection on Stroop task will be investigated, by expanding the proposed model of C-BG-TH loop to include cortico-cortico loop proposed to model attention and error detection. Even though, failure in Stroop task is mostly considered as a dysfunction of mesocortical dopamine system [Cohen D., & Servan-Schreiber D., (1992)] effecting selective attention, in this work it will be shown that failure in action selection due to dysfunction of nigrostriatal dopamine system causes similar results as prolonged time, and errors during Stroop task. Thus the proposed computational model fulfils the aim of providing a prospective look, which helps to see not yet thought relations and explanations.

A brief review of basal ganglia-thalamacortical circuits will be given in the second section. The proposed computational model of C-BG-TH loop is inspired from the works of K. Gurney et.al. and N.R. Taylor et.al.[Gurney K., Prescott T.J., & Redgrave P., (2001a), Taylor N. R., & Taylor J.G., (2000)]. In the third section first dynamical behaviour of a simple non-linear structure will be introduced then based on this structure a model of C-BG-TH loop will be proposed. This section concludes with a model of neural substrates taking part during Stroop task. This model incorporates the proposed model of C-BG-TH loop for action selection and a newly proposed cortico-cortico loop for attention and error detection. In both sections effect of dopamine on action selection and Stroop task are revealed with simulation results. In the last section, comparison of the proposed model with similar models will be given.

## 2 Neural substrates for a model of cortex-basal ganglia-thalamus loop

The basal ganglia, which are traditionally known for their role in motor movement control, now are recognized more for their part in high order cognitive processes and motivation related acts as temporal sequence storage and generation, working memory functions, behavioural switching, reward evaluation, goal-directed behaviour, reinforcement learning.

Especially, they employ suppressing of sensory inputs to let expression of one amongst them and thus originate the occurrence of an appropriate behaviour. A computational model of basal ganglia has been given as winner-take-all in [Fukai T., (1999)] and this model is in agreement with the models given in [Kropotov J.D., & Etlinger S.C., (1999), Djurfeldt M., Ekeberg Ö., &Graybiel A.M., (2001)]. Even though there are incompatible data [Tepper, J.M., Koos T., & Wilson C.J., (2004)], basal ganglia are mostly considered realizing winner-take-all function.

In order to obtain a biologically plausible computational model of basal ganglia revealing the modulatory effect of dopamine, their interrelations with related neural structures as cortex and thalamus also have to be considered. The basal ganglia-thalamocortical circuits are composed of neural structures that are responsible in generating different behavioural functions, and it has been proposed that different substructures of each neural structure take part in motor circuit, oculomotor circuit, prefrontal circuit, temporal association cortex circuit, and limbic circuit [Alexander, G.E., & Crutcher M.D., (1990a), Alexander, G.E., & Crutcher M.D., DeLong M.R., (1990b)]. On the other hand, especially mesolimbic, mesocortical and nigrostriatal dopamine systems provide regulation of information transfer through these neural circuits [Sagvolden, T., Johansen, E.B., Aase, H.& Russel, V.A., (2005)] and dysfunctions of these dopamine systems cause deficits in behavioural functions.

Our intention is to focus on a specific cognitive process, action selection, rather than modelling all aspects of a specific circuit dedicated to a specific behavioural function. So, the proposed computational model will be based on neural substrates necessary for action selection and specific substructures related with a specific behaviour as described in [Alexander, G.E., & Crutcher M.D., (1990a), Alexander, G.E., & Crutcher M.D., DeLong M.R., (1990b)] will not be considered. Thus while considering the neural substrates needed for action selection, a general approach will be followed and basic neural substrates for cortex-basal ganglia-thalamus loop will be considered rather than considering neural substructures needed to model each basal ganglia-thalamocortical circuit. Of course, the proposed model also has to reveal the effect of dopamine on action selection. Again the approach will be to model the modulation role of dopamine rather than modelling all aspects of dopamine systems and this role will be demonstrated as the effect of excess and depletion in dopamine level.

Action selection, like many other cognitive processes, is initiated at cortex where anterior cingulate system responsible for attention takes part in generating salience signal and is again terminated at the cortex where motor circuits trigger the action. The salience signal causes activation in basal ganglia, which then initiates relevant structures in cortex for action to take place via thalamus. Thus a feedback structure, which incorporates cortex, basal ganglia and thalamus, is necessary and the cortex, basal ganglia, thalamus loop provides such a structure. The generally agreed connections of basal ganglia with cortex and thalamus besides the connections within basal ganglia are depicted in Fig.1.

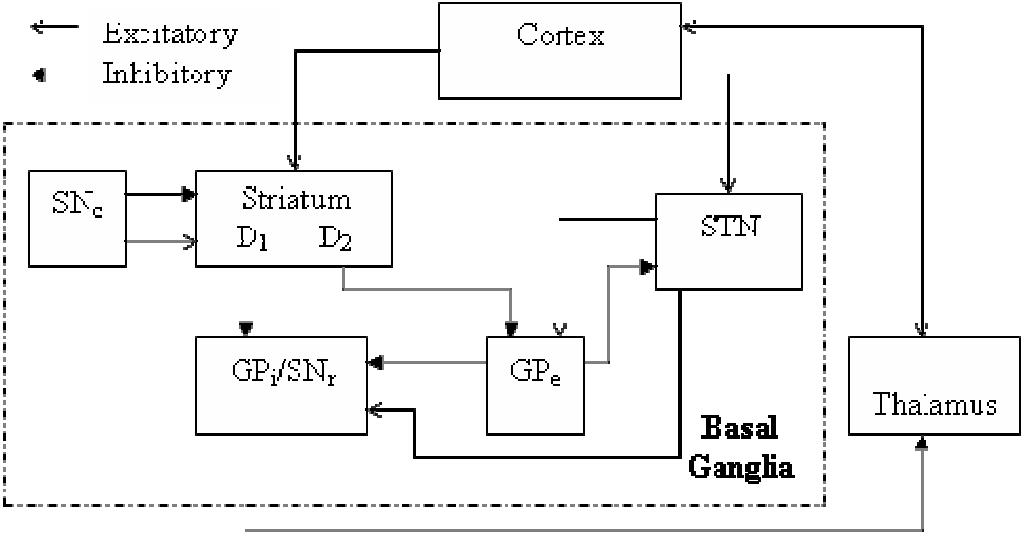

Figure 1: Cortex-basal ganglia-thalamus loop

Striatum (STR) and subthalamic nucleus (STN) are initiated by the excitatory cortical connections, which are the inputs of basal ganglia, whereas the main outputs are driven from globus pallidus internia (GPi) and substantia nigra pars reticulata (SNr). There are two main pathways in basal ganglia, information is transferred in the direct pathway through striatum and GPi / SNr and in the indirect pathway through globus pallidus externia (GPe), STN and then to GPi / SNr. While mainly GABAergic and glutamatergic projections convey this information transfer, dopamine modulates this process [Cohen J.D., Braver, T.S. & Brown J.W., (2002)].

It has been proposed that two pathways have opposing effects on thalamus, while the direct pathway disinhibits the thalamus resulting in excitation of cortex, indirect pathway has the opposite effect and inhibits the cortex [Alexander, G.E., & Crutcher M.D., DeLong M.R., (1990b), Sagvolden, T., Johansen, E.B., Aase, H.& Russel, V.A., (2005), Graybiel A.M. (1995)].While the dopaminergic signal increases the activity in direct pathway via $D_1$ receptors and disinhibiting cortex, it decreases the activity in indirect pathway via $D_2$ receptors which has inhibition effect on thalamus, thus on the overall the effect of dopamine is to let an action take place [Ouchi, Y., Kanno T., Okada H., Yoshikawa E., Futatsubashi M., Nobezawa S., Torizuka T., Tanaka K., (2001), Bar-Gad, I., Morris, G., Bergman, H., (2003) ].

In some of the computational models of C-BG-TH loop only direct pathway is considered [Amos, A., (2000), Taylor J.G., & Taylor N.R., (2000)]. However, there are also computational models where more complex interconnections are considered [Taylor N. R., & Taylor J.G., (2000), Gurney K., Prescott T.J., & Redgrave P., (2001a), Gurney K.N., Humphries M., Wood R., Prescott T.J., & Redgrave P., (2004)]. In Gurney et.al. [Gurney K., Prescott T.J., & Redgrave P., (2001a)], they argued that their intention is not to have a computational model of direct and indirect pathways but to implement a control mechanism on action selection, so their computational model includes selection pathway and control pathway rather than direct and indirect pathways.

In this work, the proposed computational model will include the direct pathway along with a modified model of the indirect pathway. The connections considered in the proposed computational model are given in Fig. 2. Even though not all the connections of indirect pathway are considered, and the effect of $D_2$ receptor is ignored still the computational model conveys the relations of the neural substrates in C-BG-TH loop. In the next section, it will be shown that the loop structure given in Fig. 2, composed of single cell for each substructure is capable of action enabling and for action selection more than one such structure is needed. Here it must be noted that single cell just models an ensemble of neurons and since the model is at behavioural level and is not one to one replication of the biological system, this simplification is valid.

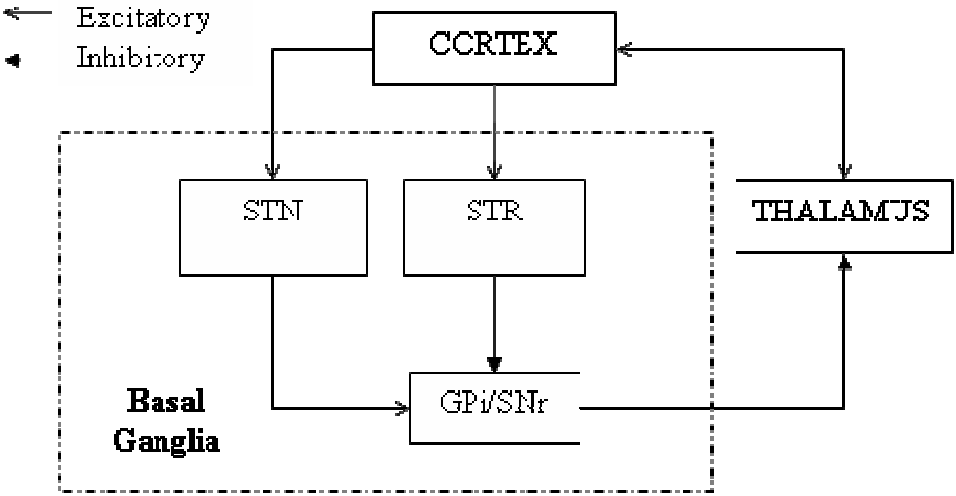

Figure 2. The neural substrates considered in the proposed computational model.

# 3 The proposed model of cortex-basal ganglia-thalamus loop

In order to establish the computational model of cortex-basal ganglia-thalamus loop introduced in the previous section, first a non-linear basic system with feedback will be defined and its analysis will be given. Based on this basic system, a model of BG-TH-C loop will be given and it will be shown that such a system is efficient in modelling action enabling. Then, how this model of BG-TH-C loop could be used to form an aggregated system for action selection will be explained. Analysis of the basic system will ease the understanding of the interpretations set up for the behaviour of the non-linear system with feedback in order to model the cognitive processes of action enabling and action selection.

## 3.1 Dynamics of the basic nonlinear system

As mentioned in section 2 each BG-TH-C loop is thought to be responsible for enabling a particular action. In order to investigate the dynamical properties of the loop, first, analysis of a non-linear basic system with feedback will be given. A discrete time dynamical system is obtained when feedback is applied to a simple cell responsible for any process and it can be represented by the difference equation,

$$x(k+1) = f(x(k), \theta, a) + I(k), \; x(0) = 0, \; I(k) = \delta(k), \; k = 0,1,2,... \qquad (1)$$

where $f(x, \theta, a) = 0.5 \cdot (1 + \tanh(a \cdot (x + \theta - 1.5)))$ and $I$ stands for the input, namely, the salience signal, which represents the importance of the corresponding action and is thought to be generated by cortex as activity in basal ganglia, is mostly initiated at cortical levels [Alexander, G.E., & Crutcher M.D., DeLong M.R., (1990b)]. Since we assume that the input is applied only at the beginning of the process, it is represented by δ(k), where δ(k) is the Kronecker delta. Using the approximation $f(0, \theta, a) \cong 0$, the system (1) becomes as follows:

$$\Sigma_0: x(k+1) = f(x(k), \theta, a), \ x(1) = I, \ k = 1,2,... \tag{2}$$

Now this system begins at $k = 1$ and the salience signal is considered as initial condition. From now on we will use this manipulation without mentioning it again.

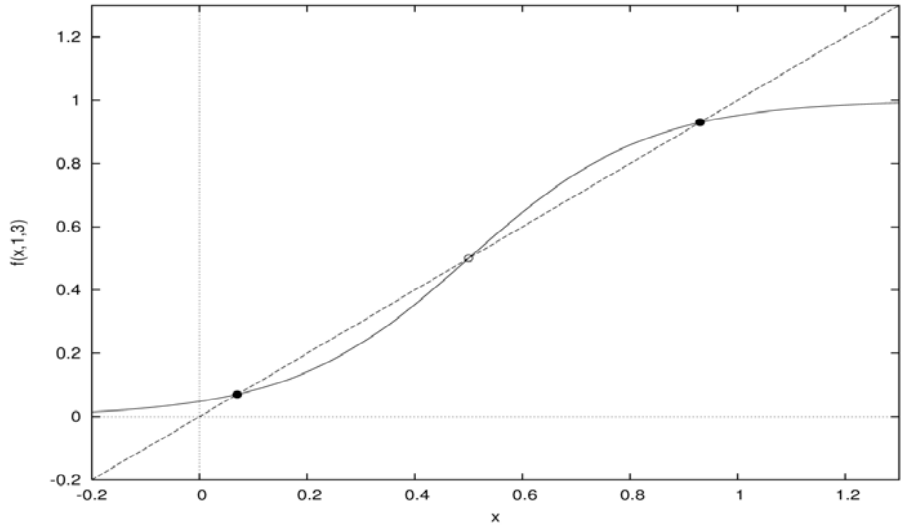

Figure 3: The activation function for the basic system with $\theta = 1$ and $a = 3$.
Filled and empty dots represent stable and unstable fixed points, respectively.

As can be followed from Fig. 3, $\Sigma_0$ has at most three fixed points; one unstable and two stable points. As the parameters $\theta$ and $a$ vary the number of stable fixed points change, which can also be followed from the bifurcation diagrams given in Fig. 4. A stable fixed point with high value is named "active point", whereas the stable point with small value is named "passive point". Excluding the unstable fixed point all initial values $x(1) = I$ of $\Sigma_0$ will drive the system to one of these stable fixed points. Hence there exist two types of attraction domains. The set of points, which constrain $\Sigma_0$ to an "active point" is called "active domain" and the other set of points is named "passive domain". According to the variation of parameter, different types of bifurcation take place. Saddle-node bifurcation occurs when $\theta$ increases, thus the number of stable fixed points increases to two. When $\theta$ is further increased, an inverse saddle-node bifurcation shows up, causing a decrease in the number of stable fixed points. For these very large $\theta$'s only the "active point" remains whereas for very small $\theta$'s "passive point" remains.

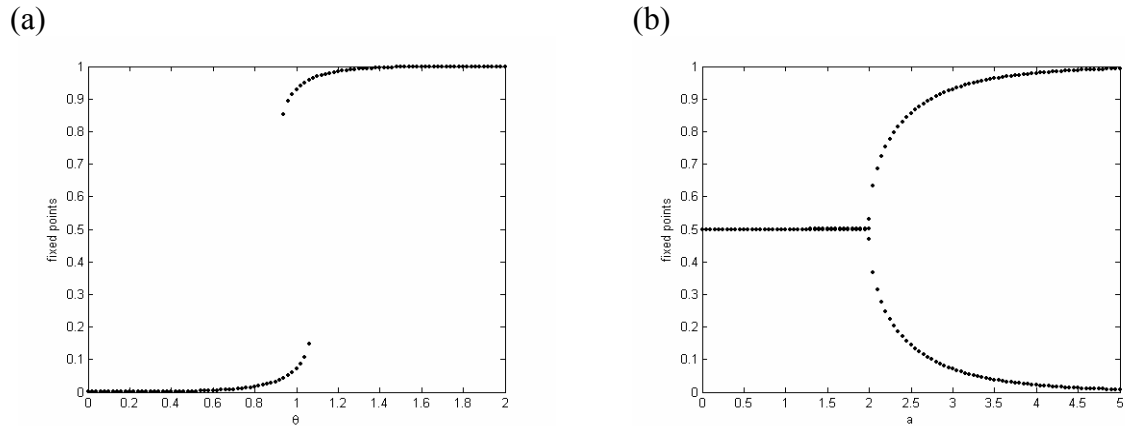

(a) (b)

Figure 4: Bifurcation diagram of $\Sigma_0$ with respect to (a) $\theta$ when $a = 3$ (b) $a$ when $\theta = 0.5$

In Fig 4(a) bifurcation diagram of the system $\Sigma_0$ according to the parameter $\theta$ is shown. Note that for intermediate $\theta$ values two stable fixed points coexist. Changing the parameter $a$ results in a pitchfork bifurcation illustrated in the Fig. 4(b). In this case one stable fixed point at 0.5 bifurcates in two stable fixed points and they get further apart as $a$ increases. The difference between the effects of $\theta$ and $a$ is that the former originates or destroys fixed points of type "active" or "passive" while the other enlarges the distance between the fixed points of $\Sigma_0$. So with parameter $\theta$, it is possible to have "active" and "passive" points coexisting besides having the possibility of only one appearing, whereas with parameter $a$ either "active" and "passive" points coexist or another stable fixed point at 0.5 exists.

## 3.2 A mathematical model for action enabling

For modelling action enabling the substructures expressed in Fig. 2 are used. While the direct pathway, which disinhibits thalamus, is completely modelled; in the indirect pathway only a single excitatory connection from STN to GPi/SNr is considered. This simplified structure will be shown to be sufficient for modelling the action enabling. Using one cell for each substructure given in Fig. 2, a model for action enabling can be constructed as in Eq. 3.

$$\begin{bmatrix} p(k+1) \\ m(k+1) \\ r(k+1) \\ n(k+1) \\ d(k+1) \end{bmatrix} = \begin{bmatrix} \lambda & 0 & 0 & 0 & 0 \\ 0 & 0 & 0 & 0 & 0 \\ 0 & 0 & 0 & 0 & 0 \\ 0 & 0 & 0 & 0 & 0 \\ 0 & 0 & 0 & 0 & 0 \end{bmatrix} \cdot \begin{bmatrix} p(k) \\ m(k) \\ r(k) \\ n(k) \\ d(k) \end{bmatrix} + \begin{bmatrix} 0 & 1 & 0 & 0 & 0 \\ 1 & 0 & 0 & 0 & -1 \\ 1 & 0 & 0 & 0 & 0 \\ 1 & 0 & 0 & 0 & 0 \\ 0 & 0 & -1 & 1 & 0 \end{bmatrix} \cdot \begin{bmatrix} f(p(k),\theta_p,a_p) \\ f(m(k),\theta_m,a_m) \\ f(r(k),\theta_r,a_r) \\ f(n(k),\theta_n,a_n) \\ f(d(k),\theta_d,a_d) \end{bmatrix} \quad (3)$$

The subsystem given by the Eq. 3 is rewritten in compact form as follows:

$$\Sigma_1: x(k+1) = \Gamma \cdot x(k) + \Lambda \cdot F(x(k),\theta,a) \quad (4)$$

In [Karabacak&Sengor], it is proved that for $\lambda \in (0,1)$ solutions of $\Sigma_1$ are bounded. Thus here, $\lambda$, standing for the inner activation of cortex, is set to 0.5. Cortex, thalamus and the substructures in BG, namely, STR, STN and GPi/SNr are denoted by $p(k)$, $m(k)$, $r(k)$, $n(k)$, and $d(k)$, respectively and all these form $x$. The initial condition has five components, $p(1)$, $m(1)$, $r(1)$, $n(1)$, $d(1)$, but they are set to zero except the value of the one corresponding to cortex, since the salience signal is thought to be generated at cortex. The importance of the first variable is, it is related with cortex and as explained in section 2, the cognitive processes are thought to initiated and terminated at cortex. So considering the first component is sufficient to decide about an action. Enabling or disabling an action could be represented by fixed points, which have small and large values in their first variables, respectively. Simulation results related to $\Sigma_1$ show that there exist two fixed points, one is approximately $(1.9\ 0.9\ 1\ 1\ 0)^T$ and the other is approximately $(0.1\ 0\ 0.1\ 0.1\ 0)^T$. Again, it is appropriate to name these "active" and "passive", respectively, just considering the value of their first components. There also exist "active" and "passive" domains, which can be defined similarly as in previous subsection. With salience value included in "active domain" $\Sigma_1$ will reach the "active point" and in this case the action is said "to be enabled" or "to be generated". Smaller salience values, which are in the "passive domain", will deprive the system to its "passive point" and this case can be interpreted as "the action is not generated".

It should be noted that for a normal action enabling both domains should coexist. In this case depending on the salience value, $\Sigma_1$ will end at "active" or "passive" point thus capable of generating or not generating an action, respectively. If only one domain exists then the system will end at the point related with the existing domain whatever the salience value is.

A set of parameter values for normal action enabling is when all $\theta$'s are set to 1 and all $a$'s are set to 3 in Eq. 3, with these parameter values both "active" and "passive" domains exist for each activation function in Eq. 3. The parameter $\theta_r$ implies the effect of neurotransmitter dopamine and to convey the effect of dopamine this parameter will be varied while keeping other $\theta$ parameters fixed at 1. Thus, excess/depletion of dopamine is simulated by an increase/decrease in $\theta_r$ value. Changing parameter $\theta_r$ will cause abnormalities in action enabling, since its effect on $\Sigma_1$ is similar to that of the basic system. The attraction domains of $\Sigma_1$ for different $\theta_r$ values are illustrated in Fig. 5. For $\theta_r$ values smaller than the normal action enabling value, there will be only "passive domain" thus whatever the salience value no action will be generated, for larger $\theta_r$ values an action will be generated even for small salience values.

As the parameter $\theta_r$ conveys the existence of either "active" point or "passive" point it is preferable over parameter $a_r$, which gives either one stable fixed point or two stable fixed points eventuated synchronously. With parameter $a_r$, it is possible to obtain normal action enabling behaviour but to model behavioural abnormalities is not possible. Though, the value of $a_r$ is still important as it determines the nonlinearity. Choosing very small $a_r$ is not appropriate because the system behaviour will be very much like a linear system and throughout this work $a_r$ will be fixed at 3 like others.

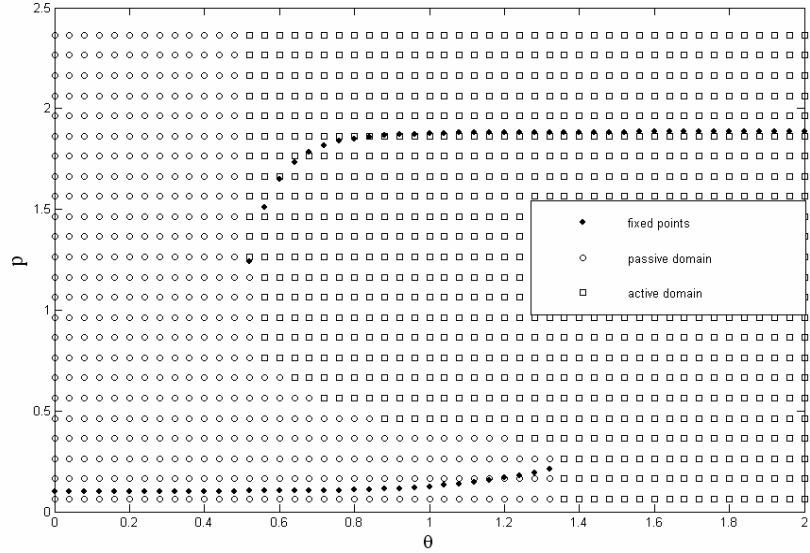

Figure 5: Domains of attraction for different $\theta$'s ($a = 3$)

## 3.3 A mathematical model for action selection

Action selection depends on competition, since in a way different neural systems had to compete amongst themselves to express behaviour or to use a common source [Gurney K., Prescott T.J., & Redgrave P., (2001a)]. How system $\Sigma_1$ executes action enabling has been investigated in subsection 3.2 and the system for action selection will be based on system $\Sigma_1$. The main idea for modelling action selection is coupling more than one $\Sigma_1$ systems in a competitive way. The coupling is biologically plausible as the diffusive, excitatory connections from STN to GPi/SNr whose overall effect is disinhibition of the other loops is used (see Fig. 6) [Gurney K., Prescott T.J., & Redgrave P., (2001a)]. The proposed model for action selection is obtained connecting two of subsystems $\Sigma_1$ as follows.

$$\Sigma_2: \begin{bmatrix} x_1(k+1) \\ x_2(k+1) \end{bmatrix} = \begin{bmatrix} \Gamma & 0 \\ 0 & \Gamma \end{bmatrix} \cdot \begin{bmatrix} x_1(k) \\ x_2(k) \end{bmatrix} + \begin{bmatrix} \Lambda & \Pi \\ \Pi & \Lambda \end{bmatrix} \cdot \begin{bmatrix} F(x_1(k),\theta) \\ F(x_2(k),\theta) \end{bmatrix} \quad (5)$$

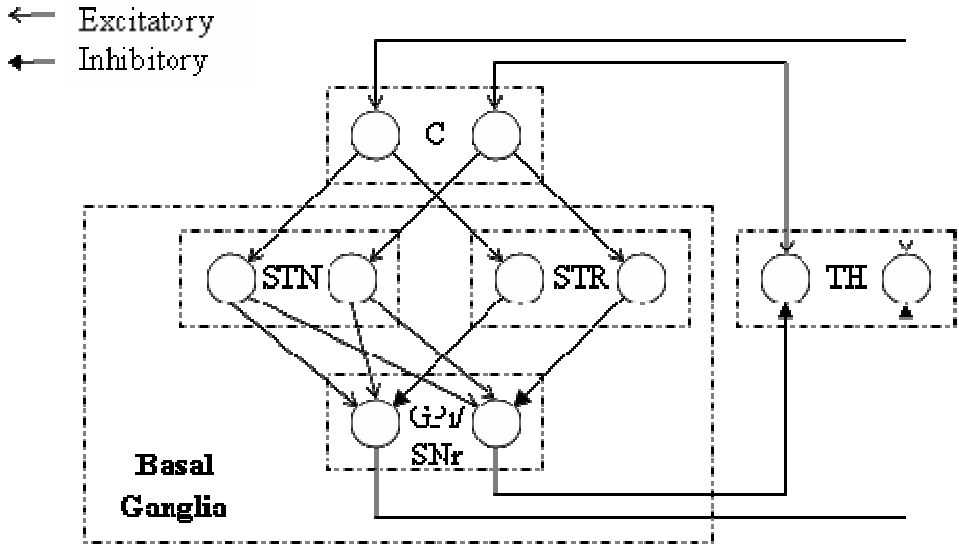

Figure 6: Model for action selection

Here, $\Pi \in \Re^{5\times 5}$ and its elements are zero except the one on fifth row fourth column, which is set to 0.5. Combining two loops, the maximum number of the fixed points increases from 2 to 4 as shown in Fig. 7. These fixed points are denoted by $x_1^*$, $x_2^*$, $x_3^*$, and $x_4^*$ and these respectively correspond to the following behaviours: both subsystems are passive, only first subsystem is active, only second subsystem is active and both subsystems are active.

For each fixed point, there corresponds a domain of attraction and these are illustrated in Fig. 7 where A, B, C and D denote the domains of attraction of $x_1^*$, $x_2^*$, $x_3^*$, and $x_4^*$, respectively. As can be followed from Fig. 7, for different initial conditions the system $\Sigma_2$ converges to different fixed points. For example, in the case $\theta = 0.7$, if the initial condition is $p_1 = 0.5$, $p_2 = 1$, $\Sigma_2$ converges to $x_3^*$ (see Fig. 7(a)). Thus the second action is selected. If the initial state of the system is in region A or D, the system will not be able to discriminate actions. Since for salience value corresponding to an initial condition in region A, none of the actions are generated and in region D both actions are generated. If for a parameter value, all of these regions A, B, C and D exist as in Fig. 7(c), the system cannot be considered as proper for action selection. For the system to act as an efficient action selector $\Sigma_2$ should have larger domains of type B and C. In Fig. 7(b) domain of attractions for parameter values that fulfils this objective is illustrated. The system with such parameter values models the normal behaviour, whereas with parameter values which create the region D and enlarge the domain A the system would model abnormal behaviours (see Fig. 7(a) and (c)). As parameter $\theta$ bears these domains of attraction, the abnormal behaviours can be interpreted to occur due to dopamine level. Similar to its effect on action enabling, change in dopamine level, i.e., dopamine excess/depletion cause both competing subsystems to get activated/inhibited for a large area of initial conditions. Thus action selection fails for these salience values and basal ganglia cannot make a proper selection.

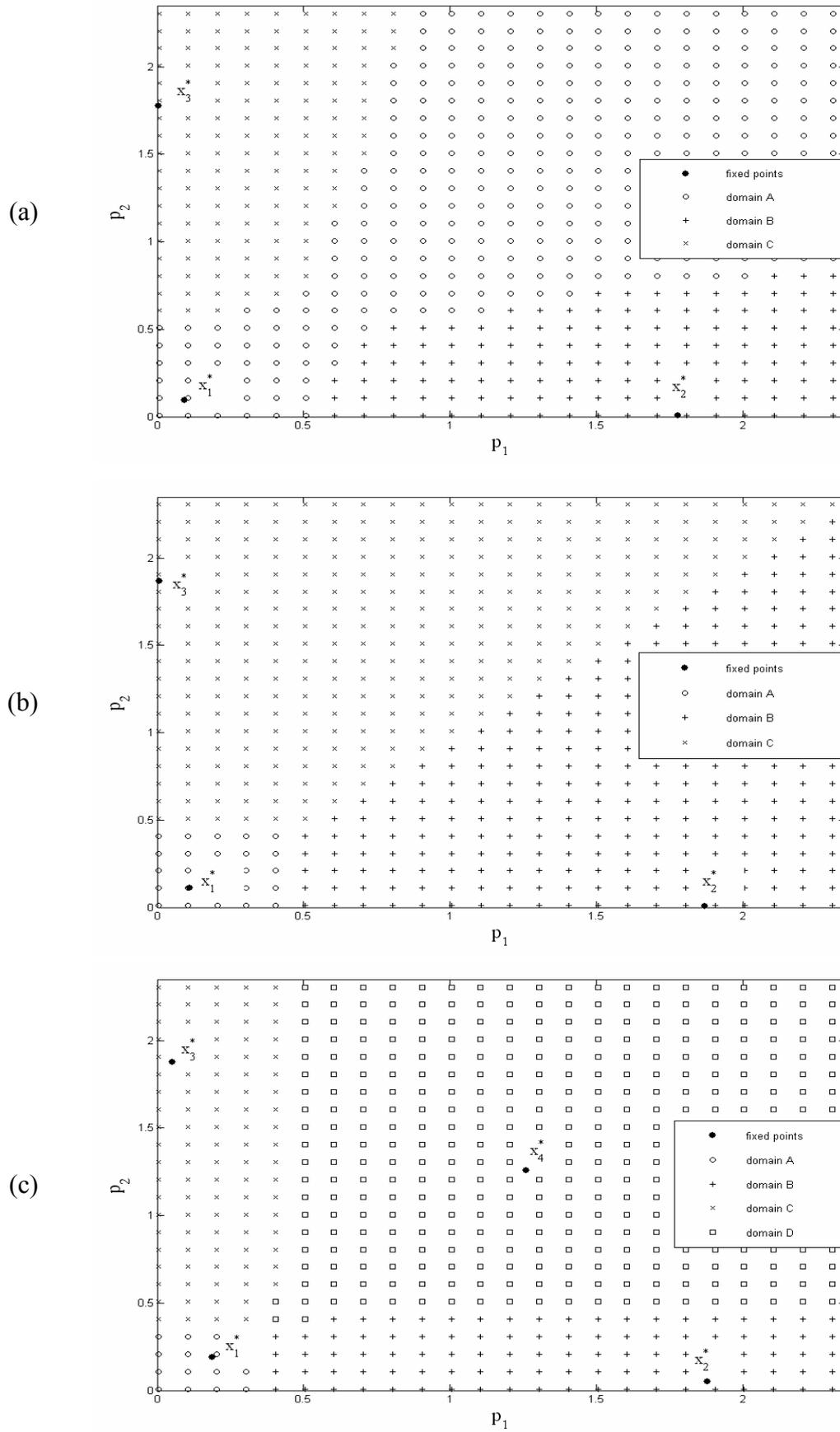

Figure 7: Attraction domains of $\Sigma_2$ for (a) $\theta = 0.7$, (b) $\theta = 1$, and (c) $\theta = 1.3$,

### 3.4 The effect of dopamine on action selection

As mentioned in the previous section, the proposed computational model of C-BG-TH loop is capable of revealing the effect of dopamine on action enabling and action selection and does this in a similar manner for both processes. In general, the effect of dopamine excess is activating. In action enabling dopamine excess forces the system to get activated even for very small salience values, whereas in action selection both actions can be selected due to the excess of dopamine. In the proposed computational model for the parameter values corresponding to the dopamine depletion level, the system for action enabling does not generate an action even for large salience values while the system for action selection does not generate either of the action. These simulation results demonstrating the effect of dopamine level on behaviours are in agreement with the literature on the interpretations of dopamine systems effect on basal ganglia [Alexander, G.E., & Crutcher M.D., DeLong M.R., (1990b), Graybiel A.M. (1995)]. Until now the basic effects of dopamine level on two cognitive functions are investigated, in this section its effect on the Stroop task will be examined.

Amongst the neuropsychological tests, Stroop task is mostly used as a measure of selective attention [Cohen J.D., Dunbar, K., & McClelland J.L., (1990), Stroop, J., (1953)]. During the task, subjects have to inhibit the word reading tendency, which is an automatic process, fast and does not require attention, and follow the novel task of colour naming, which is a controlled process, slow and requires attention [Cohen J. D., & Servan-Schreiber D., (1992)]. So, while word reading process takes less time, colour naming takes prolonged time as some time is used to inhibit the habitual behaviour. Since Stroop task is considered as a measure of focused attention, dysfunction of attentional system, i.e., anterior cingulate system is investigated in most computational models [Cohen J. D., & Servan-Schreiber D., (1992), Cohen J.D., Braver, T.S. & Brown J.W., (2002)]. Unlike previous works, in this work, effect of nigrostriatal dopamine system on Stroop task rather than mesocortical dopamine system will be investigated. This investigation has a value since the behavioural consequences of nigrostriatal dopamine system in case of occurrence of salient stimuli [Horvitz, J.C. (2000)], effect of nigrostriatal dopaminergic depletion on cognitive functions of PD (Parkinson's Disease) patients [Monchi, O., Petrides M., Doyon, J., Postuma, R.B., Worsley, K., & Dagher, (2004)] and the effect of nigrostriatal dopamine system on reinforcement learning [Djurfeldt M., Ekeberg Ö., &Graybiel A.M., (2001)] have been investigated. Thus, nigrostriatal dopaminergic system has an effect on cognitive processes in addition to its effect on motor control.

While deriving the computational model for Stroop task, the C-BG-TH loop for action selection introduced in section 3.3 is used with some alterations. In order to model the effect of attention and error detection in cortex a cortico-cortico ($C_1$-$C_2$-$C_3$) loop interacting with C-BG-TH loop is considered. Since there are two actions competing during Stroop task, namely, word reading and colour naming, two loops are constructed by originating two cells in each substructure in Fig. 8. The equations representing the dynamics of this system can be followed from Appendix. The main attribute of this system is two subsystems, the C-BG-TH loops and the $C_1$-$C_2$-$C_3$ loops working together but with different time scales. The latter runs slower, it completes one turn while the former completes its ten consecutive steps.

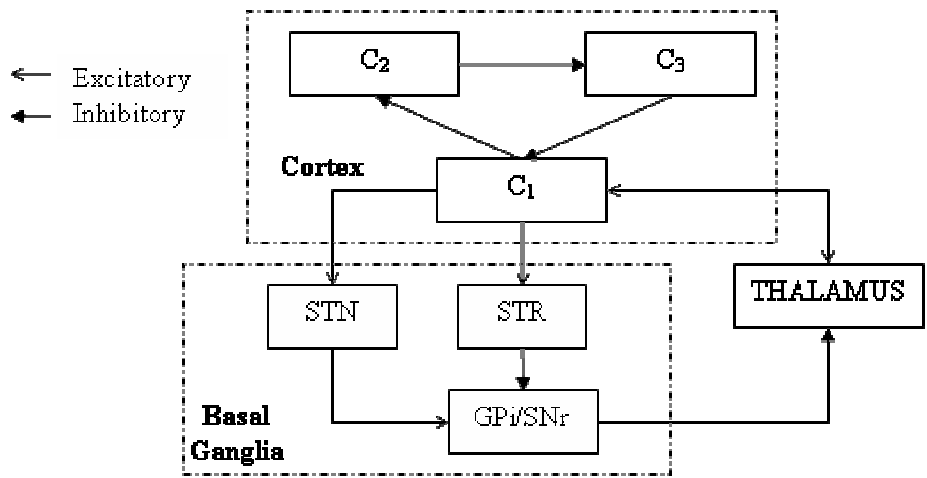

Figure 8: Model for Stroop task

Just to simulate the habit, the weight between $C_1$ and STR is increased to two in word reading loop, whereas the corresponding weight in colour naming loop remains unchanged. This causes a non-symmetric figure of attraction domains. The initial value of $C_1$ is set to $(0.5\ 0.5)^T$ representing the sensory effects of word reading and colour naming as equal due to the interference effect in Stroop task. Nevertheless, because of the abovementioned non-symmetric attraction domains colour naming loop wins even though the initial values are same. To avoid execution of the habitual action and to generate the task related one, $C_2$ and $C_3$ take their turn after ten consecutive steps of C-BG-TH loop. $C_2$, responsible for the error detection, takes the difference between the task vector $(1\ 0)^T$, and the value of $C_1$ and projects this difference to $C_3$, whose role is to focus attention on one of the actions. $C_3$ considering the output of $C_2$ affects $C_1$ by shifting its activation function on STR horizontally. Attention takes place due to shifting the activation function in the task related loop to left and in the task irrelevant loop to the right. This gives rise to activating the novel action, namely, colour naming. This can be thought as changing the non-symmetric attraction domains for the sake of the task related loop during a new 10 steps period of C-BG-TH loop. This iterative process continues till C-BG-TH loop reach its stable and probably task related fixed point.

The activation of two cells in $C_1$, which correspond to word reading and colour naming salience signals, are shown in Fig. 9 for different dopamine levels. For a normal dopamine level the loop corresponding to habitual action is activated at first, but soon the task related one takes action as can be followed from Fig. 9(b). This case can be interpreted as the colour naming task is done although at first there is a tendency to read word, which fits to the case of normal subjects performance in Stroop task and this corresponds to Stroop effect. Slight excess in dopamine results in an increase of competition period (see Fig. 9(c)). Due to dopamine excess both of C-BG-TH loops try to be activated. Thus, for a long time, signals of both actions remain at a high level until $C_1$-$C_2$-$C_3$ loops fulfil their objective, namely, to inhibit the task related action and to focus attention on the related one. This case can be regarded as the colour naming task is done but after a long waiting time. In the case of large dopamine depletion the habitual action loop is activated and later the task related one takes action but after small delay. Due to dopamine depletion it is hard to activate C-BG-TH loops, so $C_1$-$C_2$-$C_3$ loops act slower on C-BG-TH loops. Nevertheless, during first time steps habitual task loop has been activated because of the above-mentioned non-symmetric structure of attraction domains. This case can be regarded as the habitual action is done at first but then subject realizes the error and names colour.

To interpret the simulation results in the view of the previous discussion the following criteria is used: If the value of a signal is greater than one plus the other's value during 100 time steps and during the following 100 steps its value is never less than the others value minus one, then the action corresponding to this signal is regarded to be generated at the end of this 200 time steps. Using this criterion simulation results corresponding to the previous discussion are obtained and they are summarized in Table 1.

| | Dopamine Level ($\theta_{sel}$) | Error | Time of the correct response |
|---|---|---|---|
| Dopamine excess ↑ | 2 | no | > 30000 |
| | 1.6 | no | 1396 |
| | 1.4 | no | 1261 |
| | 1.2 | no | 1129 |
| | 1 | no | 396 |
| Dopamine depletion ↓ | 0.8 | no | 390 |
| | 0.6 | yes | 405 |
| | 0.55 | yes | 440 |
| | 0.5 | yes | > 30000 |

Table 1: Simulation results for different dopamine levels

The results of Table 1 depend on our criteria and interpretation of simulation results just considering the nigrostriatal dopamine system. It must be noted that other structures, especially cortical structures as sensory motor systems and phenomena as noise are also effective on occurrence of an action. These could be especially effective when dopamine excess corresponding to simulation result depicted in Fig. 9 (c) is considered. In this case prolongation of time is more than the action delay in dopamine depletion case and due to effect of unconsidered structures in the model an action and probably the wrong action could be selected beforehand.

It is well known that, depletion of dopamine in nigrostriatal system gives rise to hypokinetic disorders observed in PD, while excess of dopamine gives rise to hyperkinetic disorders observed in Huntington's Disease (HD) [Alexander, G.E., & Crutcher M.D., DeLong M.R., (1990b), Graybiel A.M. (1995)]. Even though, the results obtained for Stroop task cannot be interpreted in the view of abovementioned diseases as there is no work on Stroop task performance of PD and HD patients up to our knowledge, still the results demonstrate that deficiencies of nigrostriatal systems are effective on cognitive processes.

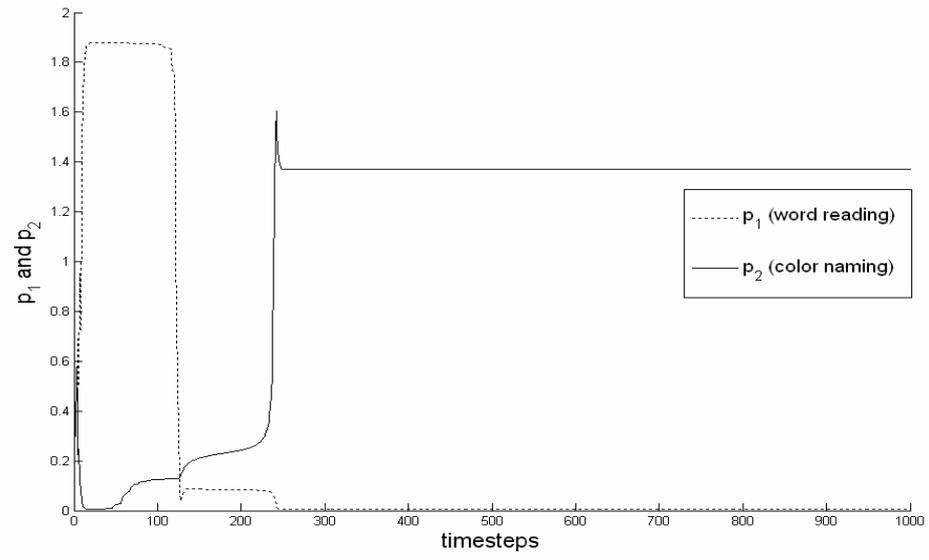

(a)

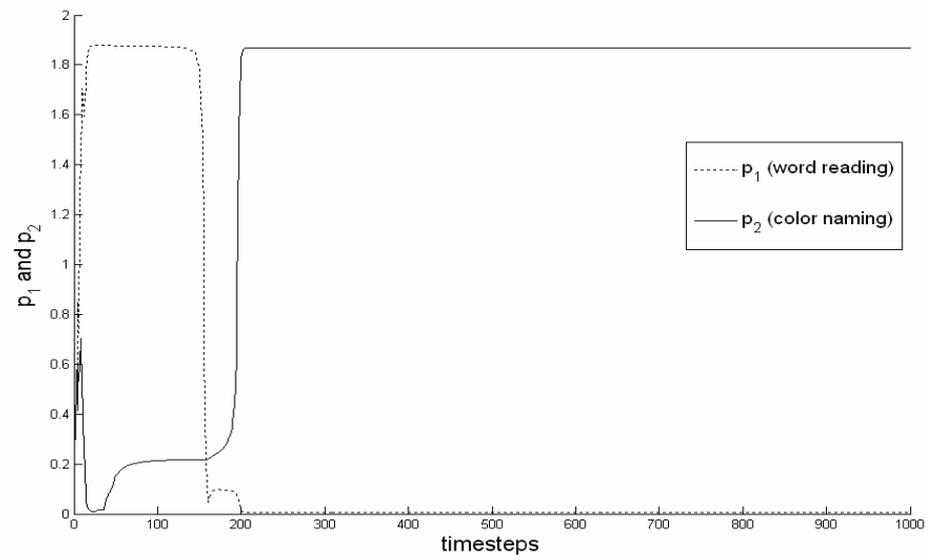

(b)

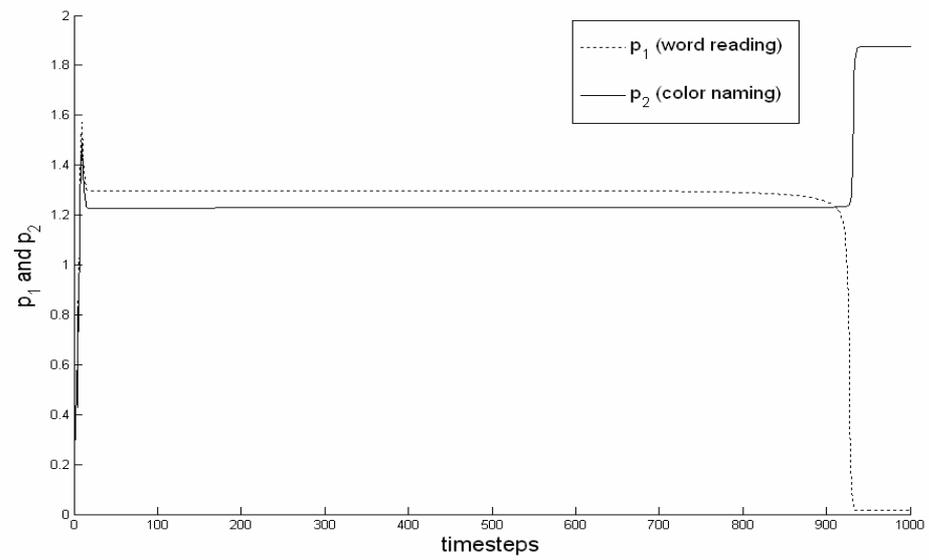

(c)

Figure 9: Activations in $C_1$ (a) for $\theta = 0.6$ (b) for $\theta = 1$ (c) for $\theta = 1.2$

## 4 Discussions and Conclusion

Quite a number of computational models of BG have been proposed and among these some evaluate the neural substrates and related neurotransmitter systems responsible for the interested behaviour. Especially models for serial processing [Taylor J.G., & Taylor N.R., (2000), Taylor N. R., & Taylor J.G., (2000)], for action selection [Gurney K., Prescott T.J., & Redgrave P., (2001a), Gurney K., Prescott T.J., & Redgrave P., (2001b)] and for reinforcement learning [Suri R.E., Bargas, J., & Arbib M.A., (2001), Doya, K., (2000)] are proposed. There are also connectionist models exploring the effect of mesocortical dopamine system on frontal cortex following similar approach [Cohen J.D.&Servan-Schreiber D., (1992), Servan-Schreiber D., Printz H., & Cohen J.D., (1990), Servan-Schreiber D., et.al., (1998), Amos A., (2000)]. Amongst these [Cohen J.D.&Servan-Schreiber D.,(1992)] considers the Stroop test. Also, there are computational models of BG dealing with physiological level rather than considering behaviour [Durstewitz D., & Seamans J.K., (2002), Ashby F.G., & Casale M.B., (2003)].

We will first focus on two of these models, Taylor J.G. et.al.'s work and Gurney K. et.al.'s work [Gurney K., Prescott T.J., & Redgrave P., (2001a), Gurney K., Prescott T.J., & Redgrave P., (2001b), Taylor J.G., & Taylor N.R., (2000), Taylor N. R., & Taylor J.G., (2000)] since these two inspired the proposed C-BG-TH loop in this work. While the proposed model of C-BG-TH loop in this work is similar to Gurney et.al.'s work for fulfilling action selection function, its rather simple dynamical equations are inspired from Taylor et.al.'s work.

A new functional architecture of BG is proposed in [Gurney K., Prescott T.J., & Redgrave P., (2001a)], where selection pathway, which is an off-center, on-surround feedforward structure performs the selection process, the control pathway regulates selection pathway through dopaminergic modulation. It has been argued that, the direct pathway and indirect pathway model of C-BG-TH loop is not sufficient in explaining all behavioural processes, even though this model is capable of explaining hypokinetic and hyperkinetic phenomena observed in PD and HD [Graybiel A.M. (1995)]. So new functional architecture of BG based on control and selection pathway may be advantageous in this respect. In Gurney et.al.'s model STN is considered as a major input of BG along with striatum and a novel role for GPe is proposed. They also investigated the effect of dopamine on action selection and interpreted the effects of dopamine depletion and excess based on the dynamical model. Considering the proposed computational model of C-BG-TH in this work, the interconnections of the two loops in Eq. 5 in Section 3.2 is inspired from Gurney et.al.'s use of diffusive interconnections and even though the dynamical model for C-BG-TH loop and modelling the effect of dopamine as a parametric shift in nonlinear activation function proposed in this work is a lot simpler than their model, similar effect of dopamine is observed.

Another work that influenced the proposed computational model is the "action network" [Taylor N.R. & Taylor J.G., (1999), Taylor J.G., & Taylor N.R., (2000), Taylor N. R., & Taylor J.G., (2000)]. The motivation of "action network" is to propose a model of C-BG-TH loop especially to model the temporal sequence storage and generation (TSSG) process, which takes part in working memory tasks. Their model is a non-linear dynamical system, and to set up the TSSG, they connected subsystems corresponding to C-BG-TH loops in layers. Each subsystem is capable of creating saddle point bifurcations due to input, which denotes the effect of other subsystems [Taylor J.G., & Taylor N.R., (2000), Taylor N. R., & Taylor J.G., (2000)]. The construction (destruction) of sustained activity is obtained by saddle point bifurcation (reverse saddle point bifurcation) and these phenomena have been interpreted as working memory. Also a discussion about HD and PD is carried and it is stated that change in a parameter denoting the effect of dopamine causes working memory dysfunction and thus reveals the effective modelling of dopamine depletion and excess [Taylor N.R. & Taylor J.G., (1999)].

The proposed model resembles that of [Taylor N.R. & Taylor J.G., (1999), Taylor J.G., & Taylor N.R., (2000), Taylor N. R., & Taylor J.G., (2000)] as it has similar simple non-linear discrete time dynamics as the model of C-BG-TH loop. As our aim is to model action selection rather than TSSG, we used diffusive parallel connections between loops corresponding to different actions. Thus the proposed model does not have a layered structure for action selection. Even though the dynamics of the proposed model is also creating saddle point bifurcation, the analysis of bifurcations due to a parameter has been done for the whole system, which helps us to reveal the overall effect of dopamine on action selection. Nevertheless, the effect of dopamine is modelled in the same way, namely, with the parameter $\theta$.

There is also the work of Servan-Schreiber D. et.al. [Servan-Schreiber D., Printz H., &Cohen D., (1990), Servan-Schreiber D., Bruno R.M., Carter, C. S., &Cohen D., (1998)] which focuses on mesocortical dopamine system. They were not concerned with C-BG-TH loop and they modelled the effect of dopamine by parameter $a$, which we have ignored after carried out the investigation in Section 3.1 which depicted that parameter $\theta$ is more convenient than parameter $a$ in modelling the effect of dopamine. This difference in chosen parameters are due to the systems considered, we modelled the C-BG-TH loop as a dynamical system. They also considered the effect of mesocortical dopamine system on Stroop task [Cohen J. D., & Servan-Schreiber D., (1992), Cohen J.D., Dunbar, K., & McClelland J.L., (1990)], and they are concerned with the effect of selective attention during Stroop task rather than action selection.

Following the work of Servan-Schreiber D. et.al., Amos A. also investigated the effect of dopamine but especially considering the performance of subjects with PD, HD and schizophrenia during WCST test. The model utilizes the idea of Servan-Schreiber so the effect of dopamine is modelled as change in gain parameter. So these models deal with the effect of mesocortical dopamine systems effect on attention circuits.

Thus based on neural substrates, in this work a computational model of C-BG-TH loop which is a nonlinear dynamical system is proposed. How the dynamics of the system is capable of action enabling and action selection is shown in detail by simulation results. These simulation results further depict the effect of nigrostriatal dopamine system on action enabling and action selection by a change of a system parameter. The proposed computational model also contributed the investigation of the effect of action selection function during the Stroop task, which is up to our knowledge, has not been considered previously. So the proposed model in a way provided a prospective look to deficits of nigrostriatal dopamine system.

# Appendix

In order to model the effect of action selection during Stroop task the system given in Eq. 3 is rewritten, where $\theta_p$ is renamed as $\theta_{att}$ and $\theta_r$ is renamed as $\theta_{sel}$ and other $\theta$'s are set to 1 as in section 3.2. This new set of equations is given in Eq. A1.

$$\begin{aligned}
p(k+1) &= \lambda \cdot p(k) + f(m(k)) \\
m(k+1) &= f(p(k)) - f(d(k)) \\
r(k+1) &= g(p(k), \theta_{att}) \\
n(k+1) &= f(p(k)) \\
d(k+1) &= -g(r(k), \theta_{sel}) + f(n(k))
\end{aligned} \quad \text{(A1)}$$

In Eq. A1 $g(x,\theta) = 0.5 \cdot (1+\tanh(3 \cdot (x+\theta-1.5)))$ and $f(x) = g(x,1)$. $\theta_{att}$ is a parameter to model the effect of mesocortical dopamine system on attention, and $\theta_{sel}$ models the effect of nigrostriatal dopamine system on action selection. The subsystem given by the Eq. A1 is rewritten in compact form considering both word reading and colour naming loops as follows:

$$\Sigma_2: \begin{aligned} \mathbf{x}_1(k+1) &= \mathbf{F}(\mathbf{x}_1(k), \theta_{att}, \theta_{sel}) + \mathbf{G}(\mathbf{x}_2(k)) \\ \mathbf{x}_2(k+1) &= \mathbf{F}(\mathbf{x}_2(k), \theta_{att}, \theta_{sel}) + \mathbf{G}(\mathbf{x}_1(k)) \end{aligned} \quad \text{(A2)}$$

The equations of $C_1$-$C_2$-$C_3$ loop are given as follows:

$$c_2(k) = \begin{cases} t - f(c_1(k-5)) &, k = 10, 20, 30, \ldots \\ c_2(k-1) &, \text{other} \end{cases}$$

$$c_3(k) = \begin{cases} A \cdot f(c_2(k-1)) &, k = 11, 21, 31, \ldots \\ c_3(k-1) &, \text{other} \end{cases} \quad \text{(A3)}$$

$$\theta_{att}(k) = \begin{cases} \theta_{att}(k-1) + 0.1 \cdot c_3(k-1) \cdot \left( 2 - \min\left( \left| \theta_{att}(k-1) - \begin{pmatrix} 1 \\ 1 \end{pmatrix} \right| \right) \right) &, k = 12, 22, 32, \ldots \\ \theta_{att}(k-1) &, \text{other} \end{cases}$$

In Eq. A3, *t* represents the input for task and set to $(0\ 1)^T$ for colour naming task. The matrix *A* is a two by two matrix with "1"'s in diagonal, "-1"'s in other entries; its role is to realize both suppression of habitual behaviour and to drive attention to novel one. The initial values are taken as very small random numbers. These two loops, namely C-BG-TH and $C_1$-$C_2$-$C_3$, are combined at $C_1$, which corresponds to the variable *p* in Eq. A1, so $c_1$ is composed of $p_1$ and $p_2$.

# References

Amos, A., (2000) "A computational model of information processing in the frontal cortex and basal ganglia", Journal of Cognitive Neuroscience, 12, 505-519.

Alexander, G.E., & Crutcher M.D., (1990a) "Functional architecture of basal ganglia circuits: neural substrates of parallel processing", Trends in Neuroscience, 13, 266-271.

Alexander, G.E., & Crutcher M.D., DeLong M.R., (1990b) "Basal ganglia-thalamacortical circuits: Parallel substrates for motor, oculomotor, "prefrontal" and "limbic" functions", Progress in Brain Research, 85, 119-146.

Ashby, F.G., & Casale, M.B., (2003) "A model of dopamine modulated cortical activation", Neural Networks, 16, 973-984.

Bar-Gad, I., Morris, G., Bergman, H., (2003) "Information processing, dimensionality reduction and reinforcement learning in the basal-ganglia", Progress in Neurobiology, 71, 439-473.

Behrendt, R. P., (2004) "A neuroanatomical model of passivity phenomena", Consciousness and Cognition, 13, 579-609.

Braver T.S., & Cohen J. D., (2000) "On the control of control: the role of dopamine in regulating prefrontal function and working memory", In Attention and Performance XVII. Control of Cognitive Processes. S. Monsell, J. Driver, Eds. 713-737.

Churchland P.S., (1988) "Neurophilosophy: Toward a unified science of the mind-brain", Cambridge, The MIT Press.

Cohen J.D., Dunbar, K., & McClelland J.L., (1990) "The control of automatic processes: A parallel distributed processing account of the Stroop effect". Psychological Review, 97, 332-361.

Cohen J. D., & Servan-Schreiber D., (1992) "Context, cortex, and dopamine: a connectionist approach to behaviour and biology in schizophrenia", Psychological Review, 99, 45-77.

Cohen J.D., Braver, T.S. & Brown J.W., (2002) "Computational perspectives on dopamine function in prefrontal cortex". Current Opinion in Neurobiology, 12, 223-229.

Cohen J.D., (2005) "The vulcanisation of the human brain: A neural perspective on interactions between cognition and emotion", Journal of Economic Perspectives, 19, 3-24.

Doya, K., (2000) "Reinforcement learning in continuous time and space", Neural Computation, 12, 219-245.

Djurfeldt M., Ekeberg Ö., &Graybiel A.M., (2001) "Cortex-basal ganglia interaction and attractor states". Neurocomputing, 38-40, 573-579.

Durstewitz, D., & Seamans, (2002) "The computational role of dopamine D1 receptors in working memory", Neural Networks, 15, 561-572.

Goel V., Pullara S.D., & Grafman J., (2001) "A computational model of frontal lobe dsyfunction: working memory and the Tower of Hanoi task", Cognitive Science, 25, 287-313.

Graybiel A.M. (1990) "Neurotransmitters and neuromodulators in the basal ganglia", Trends in Neuroscience, 13, 244-254.

Graybiel A.M. (1995) "The Basal Ganglia", Trends in Neuroscience, 18, 60-62.

Graybiel A.M. (2004) "Network-level neuroplasticity in cortico-basal ganglia pathways", Parkinonism and Related Disorders, 10, 293-296.

Grossberg, S. (1980) "How does a brain build a cognitive code?", Psychological Review, 87, 1-51.

Gurney K., Prescott T.J., & Redgrave P., (2001a) "A computational model of action selection in the basal ganglia. I. A new functional anatomy", Biological Cybernetics, 84, 401-410.

Gurney K., Prescott T.J., & Redgrave P., (2001b) "A computational model of action selection in the basal ganglia. II. Analysis and simulation of behaviour", Biological Cybernetics, 84, 411-423.

Gurney K.N., Humphries M., Wood R., Prescott T.J., & Redgrave P., (2004) "Testing computational hypotheses of brain systems function: a case study with the basal ganglia", Network: Computation in Neural Systems, 15, 263-290.

Holroyd C.B., & Coles M.G.H., (2002) "The neural basis of human error processing: Reinforcement learning, dopamine, and the error-related negativity", Psychological Review, 109, 679-709.

Horvitz, J.C., (2000) "Mesolimbocortical and nigrostriatal dopamine responses to salient non-reward events", Neuroscience, 96, 651-656.

Karabacak, O., & Sengor N.S., (2005) "A dynamical model of a cognitive function:action selection", 16th IFAC congress.

Mink, J.W., (1996) "The basal ganglia:Focused selection and inhibition of competing motor programs", Progress in Neurobiology, 50, 381-425.

Monchi, O., Petrides M., Doyon, J., Postuma, R.B., Worsley, K., & Dagher, A., (2004) "Neural Bases of Set-Shifting Deficits in Parkinson's Disease". The Journal of Neuroscience, 24, 702-710.

Norman, D. A., &Shallice T., (1986) "Attention to action: Willed and automatic control of behaviour". In Consciousness and Self-Regulation: Advances in Research and Theory. R.J. Davidson, G.E., Schwartz & D. Shapiro, Eds. Vol. 4:1-18. Plenum, New York.

Ouchi, Y., Kanno T., Okada H., Yoshikawa E., Futatsubashi M., Nobezawa S., Torizuka T., Tanaka K., (2001) " Changes in dopamine availability in the nigrostriatal and mesocortical dopaminergic systems by gait in Parkinson's disease", Brain, 124, 284-292.

Sagvolden, T., Johansen, E.B., Aase, H.& Russel, V.A., (2005) "A dynamic developmental theory of attention-deficit/hyperactivity disorder (ADHD) predominantly hyperactive/impulsive and combined subtypes", Behavioral and Brain Sciences, 28, 397-468.

Servan-Schreiber D., Printz H., &Cohen D., (1990) "A network model of catecholamine effects: gain, signal-to-noise ratio, and behaviour", Science, 249,892-895.

Servan-Schreiber D., Bruno R.M., Carter, C. S., &Cohen D., (1998) "Dopamine and the mechanisms of cognition: Part1. A neural network model predicting dopamine effects on selective attention", Biological Psychiatry, 43, 713-722.

Stroop, J., (1953) "Studies of interference in serial verbal reactions". Journal of Experimental Psychology, 18, 643-662.

Suri R.E., Bargas, J., & Arbib M.A., (2001) "Modeling functions of striatal dopamine modulation in learning and planning", Neuroscience, 103, 65-85.

Taylor N.R. & Taylor J.G., (1999) "Modelling the Frontal Lobes in Health and Disease". Proceedings of ICANN'99, London.

Taylor J.G., & Taylor N.R., (2000) "Analysis of recurrent cortico-basal ganglia-thalamic loops for working memory", Biological Cybernetics, 82, 415-432.

Taylor N. R., & Taylor J.G., (2000) " Hard-wired models of working memory and temporal sequence storage and generation", Neural Networks, 13, 201-224.